\documentclass[pra,reprint,twocolumn,showpacs,superscriptaddress]{revtex4-1}
\usepackage{graphicx}
\usepackage{graphicx}
\usepackage{amssymb,amsmath,amsthm,mathtools,empheq,cases,bm}
\usepackage{times}
\usepackage{pgfplots}
\usepackage{lipsum}
\usepackage{natbib}
\usepackage[citecolor=blue,urlcolor=blue,linkcolor=blue,colorlinks=true]{hyperref}
\usepackage{tikz}
\usepackage{pgfplots}
\usetikzlibrary{arrows,decorations.markings}
\xdefinecolor{darkgreen}{RGB}{175, 193, 36}

\usepackage[colorinlistoftodos,bordercolor=red!50,backgroundcolor=red!10,linecolor=red!50,textsize=tiny]{todonotes}
\usepackage{changes}
\setaddedmarkup{\color{red} \uline{#1}}
\setdeletedmarkup{\color{red} \sout{#1}}

\begin{document}

\title{Detecting scrambling via statistical correlations between randomized measurements on an NMR quantum simulator}

\author{Xinfang Nie}
\affiliation{Shenzhen Institute for Quantum Science and Engineering and Department of Physics, Southern University of Science and Technology, Shenzhen 518055, China}
\affiliation{CAS Key Laboratory of Microscale Magnetic Resonance and Department of Modern Physics, University of Science and Technology of China, Hefei 230026, China }

\author{Ze Zhang}
\affiliation{Shenzhen Institute for Quantum Science and Engineering and Department of Physics, Southern University of Science and Technology, Shenzhen 518055, China}

\author{Xiuzhu Zhao}
\affiliation{Shenzhen Institute for Quantum Science and Engineering and Department of Physics, Southern University of Science and Technology, Shenzhen 518055, China}

\author{Tao Xin}
\email{xint@sustech.edu.cn}
\affiliation{Shenzhen Institute for Quantum Science and Engineering and Department of Physics, Southern University of Science and Technology, Shenzhen 518055, China}
\affiliation{Center for Quantum Computing, Peng Cheng Laboratory, Shenzhen 518055, China}
\affiliation{Shenzhen Key Laboratory of Quantum Science and Engineering, Southern University of Science and Technology, Shenzhen 518055, China}

\author{Dawei Lu}
\email{ludw@sustech.edu.cn}
\affiliation{Shenzhen Institute for Quantum Science and Engineering and Department of Physics, Southern University of Science and Technology, Shenzhen 518055, China}
\affiliation{Center for Quantum Computing, Peng Cheng Laboratory, Shenzhen 518055, China}
\affiliation{Shenzhen Key Laboratory of Quantum Science and Engineering, Southern University of Science and Technology, Shenzhen 518055, China}

\author{Jun Li}
\email{lij3@sustech.edu.cn}
\affiliation{Shenzhen Institute for Quantum Science and Engineering and Department of Physics, Southern University of Science and Technology, Shenzhen 518055, China}
\affiliation{Center for Quantum Computing, Peng Cheng Laboratory, Shenzhen 518055, China}
\affiliation{Shenzhen Key Laboratory of Quantum Science and Engineering, Southern University of Science and Technology, Shenzhen 518055, China}

\date{\today}

\begin{abstract}
Out-of-time-order correlator (OTOC), been suggested as a measure of quantum information scrambling in quantum many-body systems, has received enormous attention recently.
The experimental measurement of OTOC is quite challenging. The existing theoretical protocols consist in implementing time-reversal operations or using ancillary quantum systems, therefore only a few experiments have been reported.
Recently, a new protocol to detect OTOC using statistical correlations between randomized measurements was put forward.
In this work, we detect the OTOCs of a kicked-Ising model using this new measurement method on a 4-qubit nuclear magnetic resonance quantum  simulator.
In experiment, we use random Hamiltonian evolutions to generate the random operations that are required by the randomized OTOC detection protocol.
Our experimental results are in good agreement with the theoretical predictions, thus confirming the feasibility of the protocols. 
Therefore, our work represents a step in exploring realistic quantum chaotic dynamics   in     complicated quantum systems.
\end{abstract}


\maketitle

\section{Introduction}
First introduced in the context of superconductivity~\cite{larkin1969quasiclassical,kitaev2014hidden}, out-of-time-order correlator (OTOC)
\begin{equation}\label{OTOC}
  O(t)=\langle W^\dagger (t)V^\dagger W(t)V\rangle
\end{equation}
has received much attention in both condensed matter physics and gravity physics.
Here $W$ and $V$ are  local  Hermitian operators and the expectation value $\langle \cdot \rangle$ is taken over an initial state.
This initial state can be chosen as an eigenstate~\cite{garttner2017measuring,PhysRevLett.121.016801}, e.g., the ground state,
or an arbitrary state extracted uniformly from the Haar measure to imitate the maximal mixed state at the ``infinite-temperature'' \cite{PhysRevB.96.020406}.

OTOC is recently being a key quantity to diagnose quantum chaos in condensed matter physics~\cite{hosur2016chaos,meier2017exploring}
by witnessing how fast the quantum entanglement propagates and how quantum information scrambles.
The time dependence of $O(t)$ can differentiate between regular and chaotic behavior~\cite{PhysRevX.7.031011}.
The exponential deviation of OTOC defines the Lyapnov exponent  
{$\lambda_L$} in quantum many-body systems \cite{larkin1969quasiclassical,shenker2014black}, which is upper bounded by $\beta/2\pi$, where $\beta$ is the Lyapnov exponent~\cite{Maldacena2016}.
In the context of gravity physics, it is shown that the Lyapnov exponent $\lambda_L$ saturates the bound $\beta/2\pi$
for a system which can be described holographically by an Einstein gravity~\cite{shenker2014black,shenker2014multiple,shenker2015stringy}.
Conversely, if a quantum system is exactly holographic dual to a black hole gravity, its Lyapnov exponent will saturate the bound. 
The Sachdev-Ye-Kitaev (SYK) model exhibits the fastest scrambling and saturates the upper bound of the decay rate of the OTOC,
which has been expected to shed light on the black hole information paradox~\cite{kitaev2015,PhysRevD.94.106002}.
Besides, recent theoretical studies show that OTOC   can serve as a probe of many-body localization in disordered systems with interactions~\cite{fan2017out,XChen},
and also can be used to detect both equilibrium and dynamical quantum transitions as well as the corresponding critical points in many-body quantum systems~\cite{PhysRevB.96.054503,PhysRevLett.121.016801,daug2019detection}.

The theoretical significance of OTOC functions raises the question of how to measure them experimentally.
The peculiar time order inherent in its definition makes OTOC unable to be mapped to the conventional interferometry measurement as the normal time-ordered correlators.
There are mainly two protocols proposed for measuring OTOCs, one  involving backward time evolutions and the other introducing ancillary quantum systems \cite{yao2016interferometric,PhysRevA.94.040302,PhysRevA.94.062329}.
Preciously OTOC has been measured experimentally in a nuclear magnetic resonance system~\cite{PhysRevX.7.031011,PhysRevLett.120.070501} and trapped ions~\cite{garttner2017measuring}
with the method involving backward time evolutions.
However, it remains a great challenge to implement the required time-reversal operations when dealing with many-body systems with local interactions.

In a recent study~\cite{vermersch2018}, a novel protocol to probe infinite-temperature OTOC was proposed, which requires neither time-reversal operations nor ancillary systems.
The protocol is based on statistical correlations between randomized measurements performed after time-evolution from random initial states.
The initial states are randomized with global unitary operators for the total many-body system in a global protocol.
A crucial challenge in experimental implementation of this protocol is to generate   global random unitaries efficiently.
To avoid this difficulty, Ref. \cite{vermersch2018} also devised a modified scheme, which uses local  random unitaries instead of global random unitaries.
Obviously, realizing local random operations can be significantly easier.
It is found that despite making such a simplification, the statistical correlations between the measurement results can still capture most of the real OTOC behaviour.
Summarily speaking, the key advantages of these two new protocols over the previous ones are that they neither require the reversed Hamiltonian evolution nor ancillary systems, and that they are naturally robust against depolarization and readout errors.

In this work, we study the experimental measurement of  OTOCs on a 4-qubit quantum simulator with the protocols by Ref. \cite{vermersch2018},
using techniques of nuclear magnetic resonance (NMR).
In the experiment, we  simulate the dynamics of a 4-spin kicked Ising model and probe the OTOCs over time.
The article is organized as follows.
In section ~\ref{Section2}, we briefly introduce the kicked Isinng model and the two versions of the randomized protocols.
Then the experimental realization of the two protocols are reported in section ~\ref{Section3}.
Particularly, the global random unitaries used in the global protocol  are generated with the recently proposed design Hamiltonian approach~\cite{li2018experimental,PhysRevX.7.021006}.
Finally, a conclusion and discussion is given in the end of the paper.

\section{Theory}\label{Section2}
The system we simulate in the experiment is the kicked Ising model~\cite{PhysRevX.8.021013}, i.e., an $N$-spin system with periodically driven, nearest-neighbor Hamiltonian,
\begin{equation}\label{Hkim}
\mathcal{H}(t)=\frac{J}{2}\sum_i(\sigma_i^z\sigma_{i+1}^z+h_z\sigma_i^z)+\delta(t-nT)\frac{T}{2}h_x\sum_i\sigma_i^x.
\end{equation}
Its time evolution operator of $n$ periods can be written as
\begin{equation}\label{KIM}
  U(nT)=\left[e^{-i\frac{T}{2}(J\sum_i\sigma_i^z\sigma_{i+1}^z+h_z\sigma_i^z)} e^{-i\frac{T}{2}h_x\sum_i\sigma_i^x} \right]^n,
\end{equation}
with $\sigma_i^\alpha$ $ (\alpha=x, y, z)$ being Pauli operators on the $i$-th site ($i=1, 2, \cdots, N$).
$J$ is the coupling strength between the nearest neighbors,  $h_z$ is the uniform longitudinal field, $h_x$ is the kicked transverse field,
and $T$  is the driven period.
In the experiment, we set the parameters as $h_x=J$, $h_z=0.809J$, $JT=1.6$ so that  it is a typical chaotic system.

\begin{figure}[t]
\begin{center}
\includegraphics{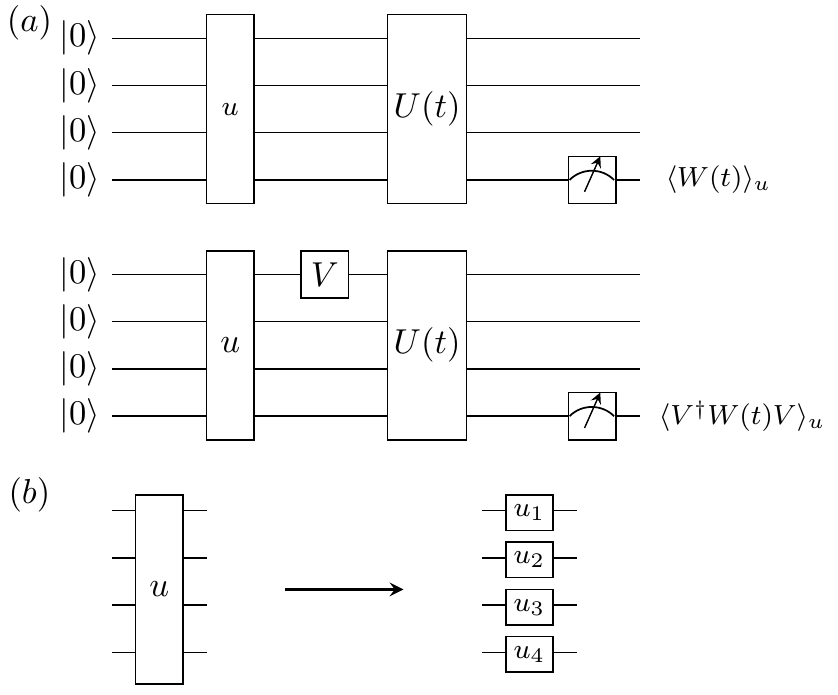}
\caption{(a) The global protocol via statistical correlations to measure OTOC $O(t)$ of a 4-spin system, in which $W$ and $V$ are local Hermitian operators on the first and fourth spins.
The protocol consists of separately random measurements of $\langle  W(t)\rangle_u$ and $\langle V^\dagger W(t) V\rangle_u$.
(c) The local random unitary operator used in the local protocol.}\label{Schemetic}
\end{center}
\end{figure}

The protocols that we intend to use for our OTOC measurement are based on statistical correlations between randomized measurements.
The randomized measurements are performed after time-evolution from randomized initial states~\cite{vermersch2018}.
Here, we  briefly introduce the global protocol for probing the  infinite-temperature  OTOCs,
in which the initial states of the measurement  is randomized with a randomized unitary $u$ sampled from the CUE~\cite{Haake} or a unitary 2-designs~\cite{PhysRevA.80.012304}.
Let $W$ and $V$ denote local Hermitian operators. Without loss of generality, we assume that they are from the Pauli matrix set. So they are  also unitary operators.
The global protocol consists of the following  three steps:

(i) Prepare the system in an arbitrary initial state $\vert \psi_0\rangle$ and apply a global randomized unitary operator $u$
to get the randomized state
$\vert \psi_u\rangle=u\vert \psi_0\rangle$.
Next, two independent experiments are performed on the same state $\vert \psi_u\rangle$:

(ii.a) In the first experiment, we evolve the system with Hamiltonian Eq.~\eqref{Hkim} and then measure the expectation value of $W$.
The circuit is   shown in Fig.~\ref{Schemetic}(a).
Steps (i) and (ii.a) are repeated for $N_t$ times with the same random unitary $u$ and different evolution time $t=nT$  ($n=1,\ldots,N_t$) to obtain $\langle W(t) \rangle_u=\langle \psi_u|U^\dagger(t)WU(t) |\psi_u\rangle$.

(ii.b) In the second experiment,  a unitary operation $V$ is applied before the evolution $U(t)$.
We also repeat steps (i) and (ii.b) for $N_t$ times  so as to get $\langle V^\dagger W(t) V\rangle_u=\langle \psi_u|V^\dagger U^\dagger(t)WU(t)V |\psi_u\rangle$,  as shown in Fig.~\ref{Schemetic}(b).

(iii) Finally,  repeat steps (i) and (ii) $N_u$ times for different sampled random unitaries $u$.

According to Ref. \cite{vermersch2018},   the  evaluation of OTOC  is closely related to the  estimation of the statistical correlation between $\langle W(t) \rangle_{u}$ and $\langle V^\dagger W(t)V \rangle_u$
\begin{equation}\label{OTOC}
  O(t)=\frac{\overline{\langle W(t) \rangle_{u} \langle V^\dagger W(t)V \rangle_u}} {\left(\overline{{\langle W(t) \rangle}_{u}^2} \overline{{\langle V^\dagger W(t)V \rangle}_u^2}\right)^{1/2}}.
\end{equation}
The notation $\overline{\cdot}$ used here denotes the average over the sampled ensemble of random unitaries.

To realize the global protocol, it is required to generate global random unitary operations from the whole unitary group.
However, one of the key  obstacles in implementing the protocol in practice is the difficulty in producing   sufficiently random unitaries.

Ref. \cite{vermersch2018} also put forward local protocol which is  experimentally easier to implement.
The local protocol starts from a product state $\vert \psi_0\rangle$ and then randomizes this state  with local random unitaries $u=u_1\otimes u_2 \cdots\otimes u_N$,
with $u_i$  a random unitary acting on the $i$-th individual spin.
It is demonstrated that the statistical correlation of the local protocol gives access to a modified OTOC $O_\text{M}(t)$,
\begin{equation}\label{OM}
  O_\text{M}(t)\equiv\frac{\sum_{A\subseteq S}\operatorname{Tr}_A(W_A(t)[V^\dagger W(t)V]_A)}{\sum_{A\subseteq S} \operatorname{Tr}_A(W_A^2(t))}.
\end{equation}
The sums in Eq.~\eqref{OM} are performed  over all subsystems $A=\{i_1,\cdots,i_{N_A}\}$ of the total system $S$, where $N_A\leq N$. The notation $(\cdot)_A$ means the reduced part of the argument in subsystem $A$, in short, $(\cdot)_A = \operatorname{Tr}_{S-A}(\cdot)$.
The modified OTOC $O_\text{M}(t)$ are sums of out-of-time-ordered functions of the different reduced operators $W_A(t)$, $VW_A(t)W$, it encodes the same physical information about scrambling with $O(t)$.
An illustration of the random unitaries used in the local protocol is shown in Fig. ~\ref{Schemetic}(c).

The key target of the experiment is to measure OTOC of the 4-spin kicked Ising model with $W=\sigma_4^z$ and $V=\sigma_1^z$.
It is analytically predicted that the long-time value of the OTOC is $O(t)=0$,
while $O_\text{M}(t)$ converges to $1/3$~\cite{vermersch2018}.

\section{Experiment}\label{Section3}
\begin{figure}[t]
\centering
\includegraphics[width=0.925\linewidth]{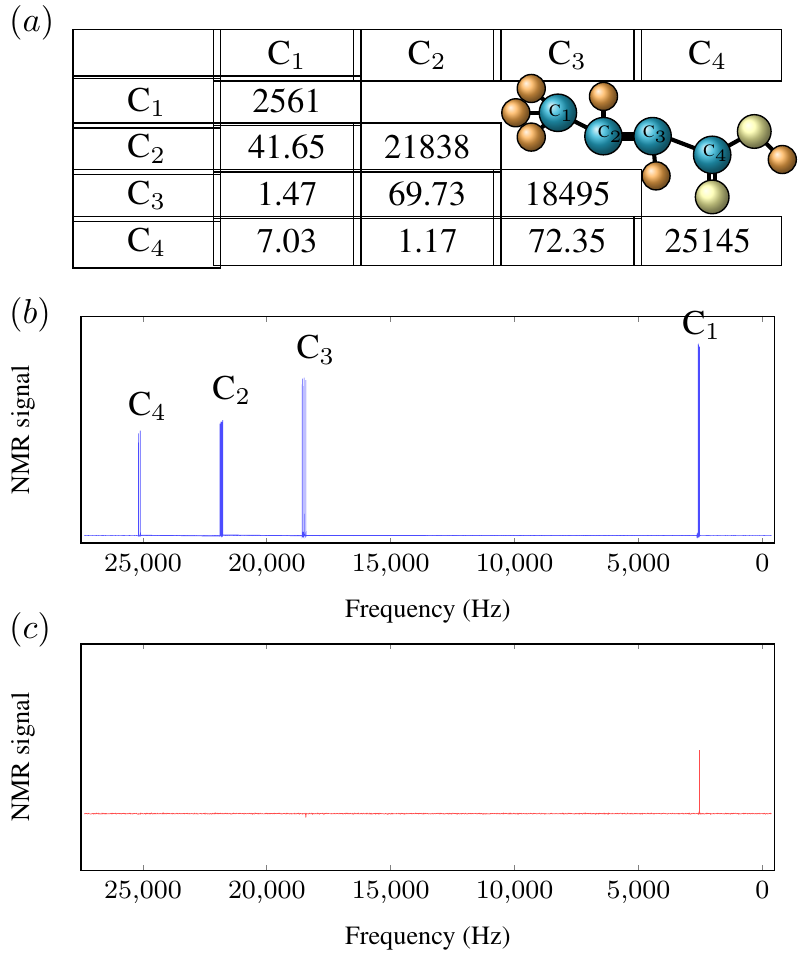}
\caption{(a) Characterization of crotonic acid. Molecular structure together with a table of the chemical shifts (on the diagonal)
and J-coupling strengths (lower off diagonal), all in Hz. The chemical shifts are given with respect to base frequency for $^{13}$C
transmitters on the $600$ MHz spectrometer that we use.
(b) Experimental spectrum of the thermal equilibrium state after a readout pulse  $\hat{\mathcal{R}}_y^1(\frac{\pi}{2})$. 
(c) Experimental spectrum of the initial state $\rho_0$ state after a readout pulse  $\hat{\mathcal{R}}_y^1(\frac{\pi}{2})$.
}
\label{Sample}
\end{figure}

We use nuclear magnetic resonance (NMR) to  simulate and probe information scrambling of the kicked Ising model.
The physical system to perform the quantum simulation
is the ensemble of nuclear spins provided by  $^{13}$C-labeled trans-crotonic acid dissolved in deuterated acetone; see Fig. ~\ref{Sample}(a) for the molecular structure.
The 4 spin-$1/2$ $^{13}$C nuclei forms a 4-qubit quantum simulator, 
and C$_1$-C$_4$ labeled in Fig. ~\ref{Sample}(a) correspond to the four spin sites in the Ising model.
The natural Hamiltonian of the  nuclear system placed in a static magnetic field along $z$-direction is
\begin{equation}\label{Hamiltonian}
\mathcal{H}_\text{NMR}=-\sum_{i=1}^4\frac{\omega_{0,i}}{2}\sigma_i^z+\pi\sum_{i<j,=1}^4\frac{J_{i,j}}{2}\sigma_i^z \sigma_i^z,
\end{equation}
where $\omega_{0,i}$ is the Larmor frequency of the $i$-th nucleus, $J_{i,j}$ is the scalar coupling between the $i$-th and $j$-th spins.
The strengths of the Larmor frequencies and the J couplings are given in Fig.~\ref{Sample}(a).
The system is controlled by the radio-frequency (RF) field, which  gives  the control Hamiltonian
\begin{equation}
  \mathcal{H}_\text{rf}(t)=\sum_{i=1}^4\frac{\omega_{1,i}(t)}{2}(\sigma_i^x\cos(\omega_\text{rf} t+\phi)+\sigma_i^y\sin(\omega_\text{rf} t+\phi)), \nonumber
\end{equation}
where $\omega_{1i}(t)$, $\omega_\text{rf}$, $\phi$ are the amplitude, frequency and the phase of the RF field, respectively.
By designing the shape of the RF pulse  delicately, we can realize any desired unitary operation on the system.
Experiment is carried out on a Bruker Ascend $600$ MHz spectrometer ($14.1$ T) at room temperature.
As illustrated in Fig.~\ref{Schemetic}, the whole experiment can be decomposed into three steps:
initial state preparation, unitary evolution and readout of the magnetization of the 4-th spin.

\subsection{Initial state preparation}
(i) Initial state preparation. This step aims to prepare the four-qubit system to the initial quantum state of quantum computation
$\rho_0=\left|0000 \right\rangle \left\langle0000\right|$.
In the high temperature limit, the nuclear system stays in the thermal equilibrium state
$\rho_{eq}\approx 1/{2^4} (I^{\otimes 4}+\sum_{i=1}^4 \varepsilon_i I_z^i)$,
where $I$ is the $2\times 2$ identity matrix and $\varepsilon_i\sim10^{-5}$ is the thermal equilibrium polarization of the $i$-th nucleus and $\varepsilon_i\propto\gamma_i$.
For there is no observable and unitary dynamical effect on the identity part, we follow the convention that we only write the traceless part of the thermal equilibrium state, i.e., $\rho_{eq}=\sum_{i=1}^4  I_z^i$.
We prepare the system from $\rho_{eq}$ to $\rho_0$ with the spatial averaging  method~\cite{10.2307/41511}.
The experimental spectra of the thermal state $\rho_\text{eq}$ and the prepared initial state $\rho_0$ are shown in Fig. ~\ref{Sample}(b) and Fig. ~\ref{Sample}(c), respctively.

\subsection{Unitary evolution}

The unitary operations involved in  the   OTOC measurment protocols described in Sec. II are the global or local random unitary $u$ and the time evolution operator $U(t)$.
To measure $\langle \sigma_1^z\sigma_4^z(t)\sigma_1^z\rangle_u$, another local gate $\sigma_1^z$ should be inserted between $u$  and $U(t)$.
Apparently,  the majority of the experimental efforts should be devoted to   pulse synthesis for implementing the random unitaries  and to simulate the desired time evolution of the periodic Hamiltonian.

\textit{Realizing random unitaries.}--
We realized both the global and local protocols in our experiments.
The random unitaries used in the local protocol were realized by choosing a set of random rotation directions and  random rotation angles for the   individual spins, and applying all the single-spin rotations simultaneously.
However, to implement the global random unitary operators  used in the global protocol requires the application of much more complicated pulse techniques.
Fortunately,  recent theoretical and experimental studies in the area of quantum pseudorandomness~\cite{PhysRevLett.116.170502,Brandao2016,PhysRevLett.116.200501,PhysRevA.97.022333,PhysRevX.7.021006,
li2018experimental,PhysRevA.96.062336,Emerson2098,PhysRevA.78.062329,harrow2009random,PhysRevLett.104.250501}
are of  great help for our present task.

Quantum pseudorandomness, which can reproduce statistical properties of random unitary operators most relevant to
quantum information tasks, plays a significant role in quantum communication and information processing.
Although quantum randomness can be effectively constructed theoretically,
realizing exact quantum randomness in many-body systems takes exponential time and is  infeasible practically.
This fact has led to the  investigation of efficient implementation of quantum pseudorandomness, particularly to the construction of unitary $k$-designs. For an introduction of k-design, see Ref.~\cite{PhysRevA.80.012304}.
There are mainly three methods to generate approximate unitary designs experimentally, i.e., local random quantum circuits composed of  single  qubit gates
and two-qubit gates acting on the nearest two  neighboring  qubits~\cite{PhysRevLett.116.170502,Brandao2016},
measurement based scheme to produce approximate designs~\cite{PhysRevLett.116.200501,PhysRevA.97.022333}
and the method of generating  random dynamics from some  design Hamiltonian~\cite{PhysRevX.7.021006}.
The first two methods are relatively harder to implement in experiment, for the random quantum circuits method requires many finely constructed local gates,
while the measurement based scheme requires plenty of physical qubits to generate the graph state.
However,  the recently proposed design Hamiltonian methodhas the merits of  not needing  local quantum gate, saving qubit resource and reducing time cost. Indeed, an experimental implementation of the design Hamiltonian approach was already achieved on an NMR system with up to 12 qubits \cite{li2018experimental}.
In our experiment, we use the same method to generate the global random unitaries in a 4-qubit system.

\begin{figure}[t]
\centering
\includegraphics[width=0.925\linewidth]{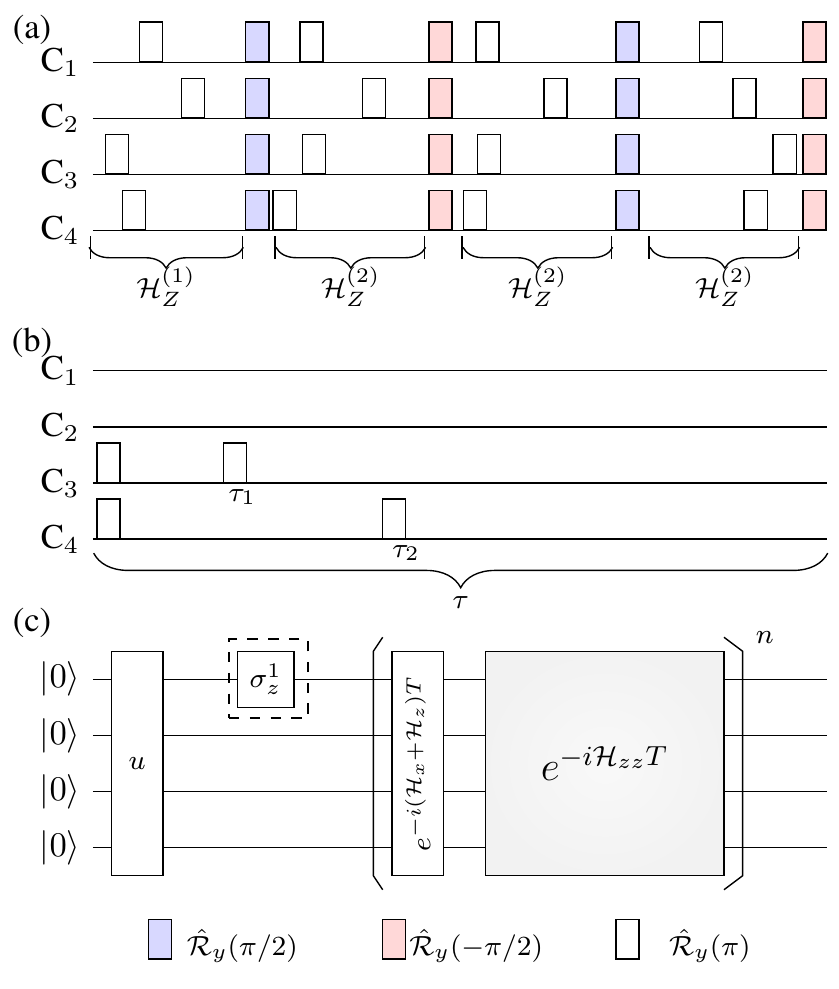}
  \caption{(a) The schematic figure about the design Hamiltonian to generate the random global unitary operators on the four-qubit system.
  For the second and the fourth random refocusing sequences,  a pair of  collective  $\pi/2$  pulses around $y$ axis or $-y$ axis are applied at the two ends of these two  sequences to to transfer $\mathcal{H}_Z^{(n)}$ to $\mathcal{H}_X^{(n)}$.
  Each  {$\mathcal{H}_Z^{(n)}$} evolves  for one half of  the period $T/2$.
  (b) Quantum circuit used to simulate the dynamics of $e^{-i\mathcal{H_\text{zz}}T}$.
  (c) The  circuit for measuring the statistical correlations between random measurements.
  The first rectangle represents the pseudorandom unitary  generated via the sequence in (a).
  $\sigma_1^z$ in the dashed rectangle means that the $\sigma_1^z$ operation is required only for the measurement of $\langle\sigma_1^z\sigma_4^z(t)\sigma_1^z\rangle$.
  The gray block represents the  quantum simulation circuit shown in  (b), aiming to   simulate the evolution operator $e^{-i\mathcal{H_\text{zz}}T}$.}.
 \label{Circuits}
\end{figure}

The experimental sequence for  generating  the 
{unitary 2-design} is shown in Fig. \ref{Circuits}(a), which consists of four random refocusing sequences
with change-of-base operations in between. 
Here, a refocusing sequence is   a sequence composed of a set of $\pi$ pulses (about the $x$ or $y$ axis). It is a common technique used in NMR spectroscopy for adjusting effective couplings between spins.
To achieve random couplings, we employ random refocusing sequences, that is,  we insert single-qubit $\pi$ pulses at random time.
Concretely, we fix the time length of each sequence to be $T/2$.  For the $m$-th refocusing sequence,  we introduce an array  $\lambda^{(m)}=\{\lambda_1^{(m)}, \lambda_2^{(m)}, \lambda_3^{(m)} , \lambda_4^{(m)}\}$   with each term  $\lambda_i^{(m)}$ draw independently  from the standard uniform distribution on the unit interval.
So each random refocusing pulse sequence consists  of four single-qubit $\pi$ pulses with  the  $i$-th  pulse applied on the $i$-th nucleus at time $\lambda_iT/2$.
The effective Hamiltonian  under the $m$-th refocusing sequence is then
\begin{equation}\label{H_eff}
  \mathcal{H}^{(m)}_Z= -\sum_{i=1}^4\omega_{i,m}^\text{eff}\sigma_i^z+\pi\sum_{i<j,=1}^4J_{ij,m}^\text{eff}\sigma_i^z \sigma_j^z,
\end{equation}
where the effective coefficients $\omega_{i,m}^\text{eff}=(1-2\lambda_i^{(m)})\omega_i$, $J_{ij,m}^\text{eff}=(1-\vert \lambda_i^{(m)}-\lambda_j ^{(m)} \vert)J_{i,j}$.
For the second and the fourth random refocusing sequence,  a pair of  collective  $\pi/2$  pulses around $y$ axis or $-y$ axis are applied at the two ends of these two  sequences,
the effective Hamiltonian will be turned into
\begin{equation}
\mathcal{H}^{(m)}_X=-\sum_{i=1}^4\omega_{i,m}^\text{eff}\sigma_i^x+\pi\sum_{i<j,=1}^4J_{ij,m}^\text{eff}\sigma_i^x \sigma_j^x.
\end{equation}
Now the Hamiltonian  under  the whole pulse sequence is
\begin{equation}
	\{\mathcal{H}_Z^{(1)}, \mathcal{H}_X^{(2)},\mathcal{H}_Z^{(3)}, \mathcal{H}_X^{(4)}\}.
	\label{designH}
\end{equation}
It is expected that through such an alternate change between the  Pauli-$Z$ and Pauli-$X$ bases  would  allow the system's time-evolution quickly approach a unitary 2-design after sufficiently long time.

\begin{figure}[t]
\begin{center}
\includegraphics[width=0.95\linewidth]{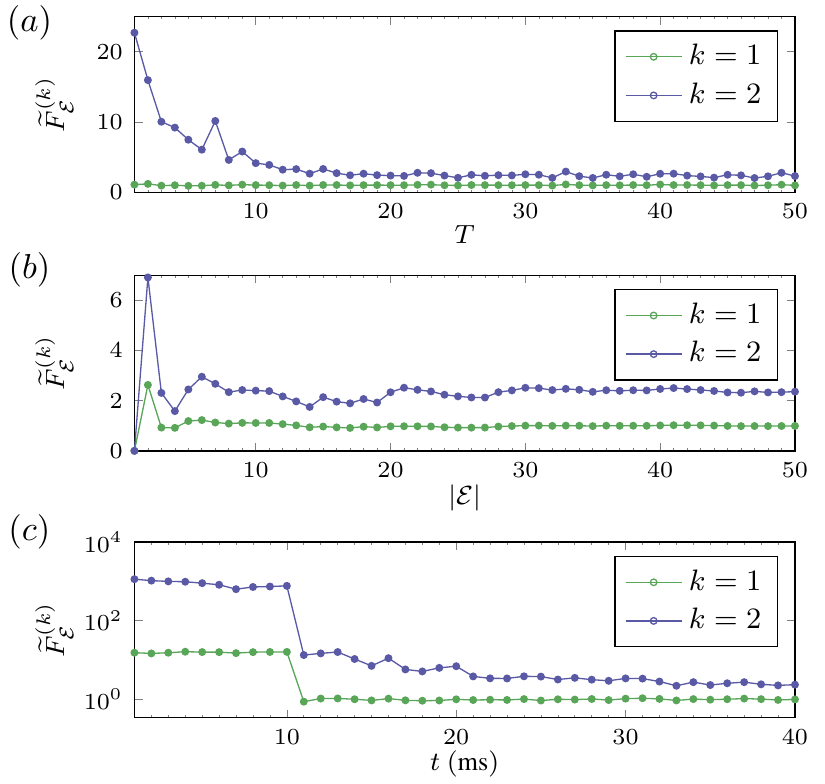}
\end{center}
\caption{The numerical simulation of the first and second  frame potential. The simulation data here are based from $50$ randomly chosen $\lambda$.
(a)-(b) The estimated frame potentials achieve the convergence of $\widetilde{F}_\mathcal{E}^{(1,2)}$ as the period prolonged and the sample size enlarged.
(c) Convergence of $\widetilde{F}_\mathcal{E}^{(1,2)}$ estimated from  the random $\lambda$. $\widetilde{F}_\mathcal{E}^{(1,2)}$  drops suddenly
at the time the change of axis operation is applied, and  gradually approach its Haar value.}\label{FP}
\end{figure}

To confirm the above statement, we compute the $k$-th frame potential $F_\mathcal{E}^{(k)}$ ~\cite{li2018experimental,10.1063/1.1737053}  for $k=1$ and $k=2$ to check whether the random Hamiltonian can  generate a unitary 2-design.
In Fig. ~\ref{FP}(a), we compute the frame potentials  $\widetilde{F}_\mathcal{E}^{(1)}$ and $\widetilde{F}_\mathcal{E}^{(2)}$ with different periodic time $T$  for $50$ random unitaries $u$  generated  from $50$  randomly sampled  $\{\lambda^{(1)},\lambda^{(2)},\lambda^{(3)},\lambda^{(4)}\}$.
Fig.~\ref{FP}(b) shows the numerical results of $F_\mathcal{E}^{(k)}$ with $T=20$ ms against the sample size $\vert \mathcal{E} \vert$.
Convergences of  $\widetilde{F}_\mathcal{E}^{(1)}$ and $\widetilde{F}_\mathcal{E}^{(2)}$ to their corresponding Haar values $k!$ with respect to $T$ and  $\vert \mathcal{E}\vert$ are  both  observed.
Fig. ~\ref{FP}(c)  presents the numerical simulation results for different evolution time $t$ of the random Hamiltonian Eq. (\ref{designH}), which suggests that for a long time of evolution, that is,  after about two rounds of evolution,
the estimated frame potentials converge to their corresponding Haar values.
The numerical  simulation results show that the random Hamiltonian is random enough to generate unitary 2-designs when the period is chosen as $T=20$ ms.

\begin{figure*}
\centering
\includegraphics[width=0.925\linewidth]{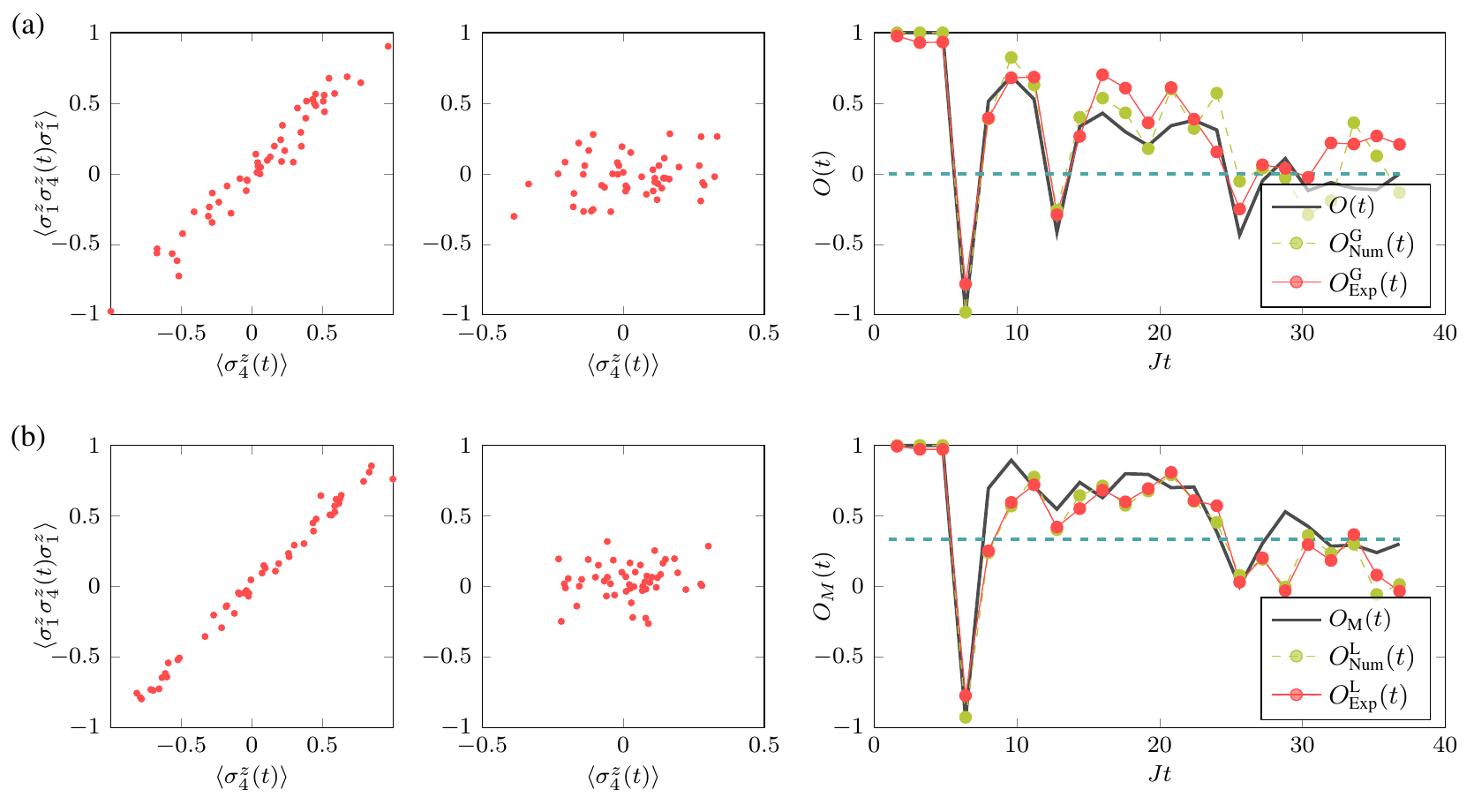}
\caption{(a) The experimental results of the global  protocol of the Kicked Ising model with $h_x=J$, $h_z=0.809J$, $JT=1.6$, $N=4$, $W=\sigma_4^z$ and $V=\sigma_1^z$.
The first two panels show the experimental statistical distributions of the measurement results obtained with $N_u=50$  at $JT=1.6$ and $JT=32$.
In the third panel, the experimentally exacted statistical correlations $O_\text{Exp}^\text{G}(t)$ (red dots) and
the numerically simulated ones $O_\text{Num}^\text{G}(t)$ (dark green dots), together with the  theoretical  exact OTOCs $O_\text{M}(t)$ (solid black line) are shown.
The dashed line represents the analytical predictions of $O(t)$.
(b) The same as (a) with local protocol, the statistical correlations $O_\text{Exp}^\text{L}(t)$, $O_\text{Num}^\text{L}(t)$ compared with the modified OTOC $O_\text{M}(t)$.
The dashed line represents the analytical predictions of $O_\text{M}(t)$.
}\label{Results}

\end{figure*}

\textit{Simulating the kicked Ising model.}--
The key part of our experiment consists in implementing the unitary evolution of the kicked Ising model Eq.~\eqref{KIM}.
We rewrite $U(T)$ as
\begin{equation}\label{Trotter}
  U(T)=e^{-i\frac{T}{2}\sum_iJ\sigma_i^z\sigma_{i+1}^z}e^{-i\frac{T}{2}h_x\sum_i\sigma_i^x}e^{-i\frac{T}{2}h_z\sum_i\sigma_i^z}.
\end{equation}
Apart from the first term, the other two terms are global rotations around $x$ or $z$ direction, which can be easily done with hard pulses.
The main problem remaining now is how to realize the first term $e^{-i\frac{T}{2}\sum_iJ\sigma_i^z\sigma_{i+1}^z}$.
We can still use the refocusing sequence technique for accomplishing this task as a suitably designed refocusing scheme will allow
equalizing the couplings between the nearest spins,
while eliminating the other unwanted couplings.
Although there exists a general and efficient refocusing scheme for any $\sigma^z \sigma^z$-coupled evolution~\cite{PhysRevA.61.042310},
it is possible to find a much simpler circuit for the current task.
Fig. ~\ref{Circuits}(b) shows our constructed  circuit.
Concretely, we do the following steps. First, we neglect the evolution effects of  $J_{13},J_{24},J_{14}$ 
for  that their values are much more smaller than those of $J_{12}, J_{23}, J_{34}$.
Two single-qubit $\pi$ pulses are applied on  C$_3$ and C$_4$ nuclei respectively to tune the effective J-couplings $J_{23}^\text{eff}$ and $J_{34}^\text{eff}$ to $J_{12}$.
The parameters of the evolution should fulfil the following requirements to yield the right evolution:
\begin{equation}
\begin{dcases}
\pi J_{12}\tau/2= JT/2, \\
(2\tau_1-\tau)J_{23}=J_{12}\tau, \\
[2(\tau_2-\tau_1)+\tau]J_{34}=J_{12}\tau.
\end{dcases}
\end{equation}
For $JT=1.6$, the solution of the equations are 
$\tau=12.23$ ms, $\tau_1=9.77$ ms and $\tau_2=7.17$ ms.
The time evolution of the chemical shifts during this duration can be eliminated with a collective single-qubit rotation around $z$ axis,
the  expression  of which can be written as $R_z=\prod_i e^{-i\alpha_i\sigma_i^z/2}$.
Here, the rotational angles satisfy the following conditions,
\begin{equation}
\begin{dcases}
\alpha_1  =\omega_1 \tau, \\
\alpha_2  =\omega_2 \tau, \\
\alpha_3 =\omega_3 (2\tau_1-\tau), \\
\alpha_4 =\omega_3 (2\tau_2-\tau).
\end{dcases}
\end{equation}
The mulitple-qubit rotation $R_z$ is implemented together with the last two terms in Eq.~\eqref{Trotter}, $e^{-i(\mathcal{H}_x+\mathcal{H}_z)T}=e^{-i T (h_x\sum_i\sigma_i^x+\sum_i(h_z+ \alpha_i / T)\sigma_i^z))/2}$, as shown in the quantum circuit of the whole experiment Fig.~\ref{Circuits}(c) to measure OTOC.

Therefore, the whole quantum network to simulate the kicked Ising model includes single-qubit rotations, global rotations and free evolutions of the natural Hamiltonian of the  nuclear spin system $\mathcal{H}_\text{NMR}$.
Each single-qubit rotation is realized with a shaped pulse with a length  of $0.5$ms.
To further improve the performance of the pulse sequence, we optimise the two unitary parts, including the pseudorandom unitary operator generation part and the quantum simulation of kicked Ising model, with the gradient ascent pulse engineering (GRAPE) technique~\cite{KHANEJA2005296}.
For the global protocol, a random unitary $u$ is realized with a single shaped pulse with a duration of $50.5$ ms.
A local random unitary used in the local protocol is realized  with a shaped pulse with length $0.5$ ms.
The evolution of the kicked Ising model $U(nT)$ is realized with a pulse with length $13.74n$ ms.

(iii) Readout.
To detect the OTOC dynamics of the kicked Ising model using statistical correlations between randomized measurements, we choose $50$ different random unitaries
and simulate the dynamics of Ising model up to $23$ time periods.
The statistical correlations are based on measuring the expectation value of 
$\sigma_4^z$.
Since $\langle \sigma^z_4\rangle=2\langle  I_z^4\rangle$, the expected OTOC is estimated through measuring the expectation value of the longitudinal magnetization
of the nucleus  C$_4$, i.e.,  $\langle I_z^4\rangle$.
The readout of the expectation value in NMR system is just an ensemble-averaged one over the ensemble of the molecules.
By applying a  single-qubit $\pi/2$ rotation along the $y$ axis applied on nucleus C$_4$, $\langle I_z^4\rangle$ is readout by measuring the NMR spectrum signal.
$\langle  \sigma_4^z(t)\rangle_u$ and $\langle  \sigma_1^z \sigma_4^z(t) \sigma_1^z\rangle_u$ correspond to the $\sigma_4^z$ of the final state
 without and with unitary operator $\sigma_1^z$ applied between $u$ and $U(nT)$,  respectively.

\section{Results}
The experimental results using the two protocols to detect the infinite-temperature OTOCs  of a 4-spin kicked Ising model are shown in Fig.  ~\ref{Results}.
The results of the global protocol are illustrated  in Fig. ~\ref{Results}(a).
The first two panels show the experimental statistical distributions after $50$ random measurements at $Jt=1$ and $Jt=32$.
Each statistical panel gives out a statistical correlation to estimate the  OTOC at the corresponding time $t=nT$.
Because $\sigma_z^1$  and $\sigma_z^4$  commute to each other at $t=0$, the value of the  OTOC at the beginning several periods is almost  unity.
In the third panel, the extracted experimental statistical correlations $O_\text{Exp}^\text{G}(t)$ (red dots) are given.
The theoretical OTOCs $O(t)$ (black  curves), numerically simulated statistical correlations $O_\text{Num}^\text{G}(t)$ (green dots) with the same sample set of random $u$
and the analytical OTOCs for the 
infinite Ising model (dashed line) are also presented for comparison.
Fig.~\ref{Results}(b) shows the experimental results of the local protocol, 
the solid black line, the green dots  and  the dashed line represent the modified OTOCs $O_\text{Num}^\text{L}(t)$, $O_\text{M}(t)$ and the analytical $O_\text{M}$, respectively .

As shown in panels Fig.~\ref{Results}(a) and Fig.~\ref{Results}(b), in both the global and local protocols, the measured  experimental evolutions of $O_\text{Exp}^\text{G}(t)$ and $O_\text{Exp}^\text{L}(t)$ resemble well
the ones of the theoretical  OTOC $O(t)$ and the modified OTOC $O_\text{M}(t)$, respectively.
The divergences between $O_\text{Num}^\text{G}(t)$ and $O(t)$ is mainly due to that the $N_u=50$ random measurements are not enough.
The same reason for local protocol.
The experimental data agree well with the numerical simulated results.
The sources of the experimental errors include imperfections in state preparation, control inaccuracy and decoherence.

\section{Conclusion and Discussions}
To conclude, we have experimentally demonstrated  both the global and  local protocols to detect the infinite-temperature OTOC using statistical correlations between randomized measurements.
As an example, we measure OTOCs of a typical chaotic system, i.e., a four-spin kicked Ising model.
Specifically, the global random unitary evolutions used in the global protocol are constructed with a design Hamiltonian method.
The design Hamiltonian technique may find more applications in  future quantum information experimental researches,
and will serve as a theoretical tool for understanding non-equilibrium dynamics of quantum many-body systems,
such as  in the studies of  quantum communication~\cite{hastings2009superadditivity,PhysRevLett.92.187901},
quantum entanglement measurement~\cite{PhysRevLett.108.110503,PhysRevLett.120.050406},
quantum chaos~\cite{PhysRevX.8.021062} and quantum thermalization~\cite{PhysRevLett.122.070601}.
Our experimental successful detection of OTOC shows the feasibility and  the  usefulness of the randomized  protocol.
The protocol can also be used to detect information scrambling in other quantum systems, 
such as  Rydberg atoms, trapped ions or superconducting qubits.
One important problem left open for future   exploration is to extend the current randomized protocol to the experimental detection of  OTOCs of finite temperature.

\emph{Acknowledgments}.
J. L., T. X. and D. L. are supported by
the National Natural Science Foundation of China (Grants
No. 11605005, No. 11875159 and No. U1801661),
Science, Technology and Innovation Commission of Shenzhen
Municipality (Grants No. ZDSYS20170303165926217
and No. JCYJ20170412152620376), Guangdong Innovative
and Entrepreneurial Research Team Program (Grant No.
2016ZT06D348).
J. L. is supported by the  National Natural Science Foundation of China (Grants  No. 11605005).


%

\end{document}